# [*]Nature of magnetism in the spin-chain compound, $Ca_3CuRuO_6$

## S. RAYAPROL AND E. V. SAMPATHKUMARAN
Tata Institute of Fundamental Research, Homi Bhabha Road, Mumbai – 400 005

*Abstract*

*A quasi one- dimensional compound, $Ca_3CuRuO_6$, has been synthesized by solid-state reaction method and studied using magnetization (M) and heat capacity (C) measurements. This compound undergoes magnetic ordering ($T_o$) around 40 K, as evidenced by the dc magnetic susceptibility ($\chi$) behavior. However, the magnitude of the paramagnetic Curie temperature ($\theta_p$) obtained from the high temperature linear region is large (**–277 K**, with the negative sign indicating antiferromagnetic interaction). The reduction of $T_o$, compared to $\theta_p$, is attributed to geometrical frustration effect arising from the triangular arrangement of antiferromagnetically coupled magnetic chains. The absence of a feature in ac $\chi$ around 40 K rules out possible spin-glass freezing. However, we find that the peak in C(T) around 40 K is weak, with the entropy change associated with the transition being negligible, typical of a disordered magnetism. We infer that this material thus exhibits inhomogeneous magnetism, despite being stoichiometric, presumably due to interplay between geometrical frustration and disorder in the Cu-Ru chain.*

## INTRODUCTION

The structure of $A_3MXO_6$ (where A = Ca or Sr and M/X=alkali or transition metal ions) type of compounds, derived from rhombohedra of $K_4CdCl_6$, consists of one-dimensional chains of alternating face sharing $MO_3$ (trigonal prisms) and $XO_3$ (octahedra) along the c-axis. This structure is characterized by the triangular arrangement of magnetic ions (M/X) in the basal plane and hence provide a nice opportunity to probe magnetic frustration effects arising out of topological reasons under favorable circumstances. Hence, these compounds have started attracting the attention of condensed matter physicists.

While most of the $A_3MXO_6$ compounds crystallize with rhombohedral symmetry in $R\bar{3}c$ space group [1], the Cu-containing compounds are of special interest due to crystallographic distortions arising out of Jahn-Teller effect, thereby causing interesting magnetic anomalies [2-5]. The Cu ions (occupying the trigonal prismatic site) are displaced towards one of the rectangular faces, thus adopting a distorted square planar coordination. This displacement is ordered throughout the structure resulting in the lowering of crystal symmetry from rhombohedral to monoclinic [2-5].

As a continuation of our efforts in this class of compounds, we present the results of bulk measurements (dc and ac $\chi$, C and high-field isothermal M) on the compound, $Ca_3CuRuO_6$ which forms in a monoclinic structure. It may be recalled that there was a previous report on the dc $\chi$ behavior of a non-stoichiometric compound, $Ca_{3.1}Cu_{0.9}RuO_6$, but the authors could not synthesize single-phase stoichiometric compound [5].

## EXPERIMENTAL DETAILS

Polycrystalline sample of the $Ca_3CuRuO_6$ was synthesized by solid-state reaction method. Appropriate quantities of the starting components - $CaCO_3$, CuO and Ru metal powder (> 99.9 % pure) - were grinded thoroughly under acetone. The samples were then calcined at 800$^0$C for 12 hours and sintered as pellets in the temperature range of 800 – 1200 $^0$C for about 7 days with intermediate grindings. The samples were characterized by x-ray diffraction (Cu-$K_\alpha$) and found to be single phase with a monoclinic distortion. The $\chi$ (both ac and dc) and isothermal M measurements were carried out with commercial magnetometers. The C data were collected by relaxation method with PPMS (Quantum Design).

## RESULTS AND DISCUSSION

Figure 1 shows the x-ray diffraction pattern for $Ca_3CuRuO_6$. There is no evidence for any extra phase reported in Ref. 5. This establishes that the stoichiometric compound can be formed. The lattice constants were determined by least squares method and the values (shown in the figure 1) are close to those reported for the non-stoichiometric composition [5].

---

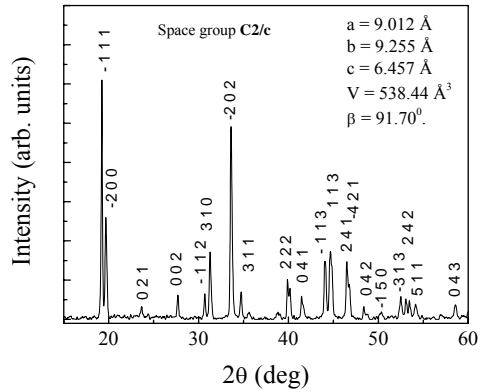

**Fig 1.** X-ray diffraction pattern of $Ca_3CuRuO_6$ (Cu-$K_\alpha$).

The $\chi^{-1}$ follows the Curie-Weiss behavior above 100K as shown by the straight line passing through the data points in figure 2. The values of $\theta_p$ and $\mu_{eff}$ obtained from the Curie-Weiss fit in the linear region are found to be  *–277 K* and *4.02 $\mu_B$/f.u* respectively. The plot of dc $\chi(T)$ (figure 2) shows a sudden upturn around 40 K, with a peak around 20 K signaling the existence of a magnetic transition. These findings are in broad agreement with those reported for non-stoichiometric compound. On the basis of neutron diffraction data on the non-stoichiometric composition, it was proposed previously [5] that a ferrimagnetic structure involving Ru and Cu magnetic moments oriented parallel to the basal plane sets in around 40 K. If so, one would naively expect a feature in the ac $\chi$ as well around 40 K. We have therefore performed such studies, and, to our surprise, there is no feature either in the real as well as in the imaginary parts of ac $\chi$, thus ruling out any long-range magnetic ordering, and also spin-glass-like freezing. The question therefore remains what the nature of this magnetic transition is.

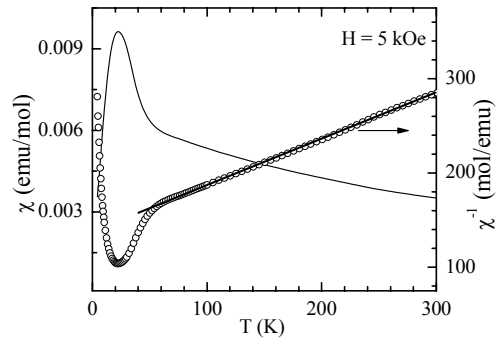

**Fig 2** Dc magnetic susceptibility for $Ca_3CuRuO_6$.

In order to address this issue further, we have performed C measurements in the range 2 – 100 K (see figure 3). It is to be noted that the feature around 40 K is extremely weak both in C vs. T as well as C/T vs. T plots. Clearly, the entropy (S) associated with the transition (Fig. 3) is negligible. Absence of a prominent feature in C(T) is not typical of a long-range magnetic ordering, but of an inhomogeneous magnetism. Spin-glass freezing cannot be the origin of this inhomogeneous magnetism as inferred from ac $\chi$ behavior discussed above.

One way to consistently understand all the above results, including possible ferrimagnetism proposed from neutron data on the non-stoichiometric compound [5], is to propose that a small fraction of Ca ions are present along the chains (due to chemical disorder). This defect makes the magnetic chain lengths finite and random in length, and the 'ferrimagnetism' sensed by neutron diffraction data is of a cluster type (nanomagnets?). Possible presence of such defect-induced random-sized clusters is presumably responsible for the 'inhomogeneous magnetism' inferred from the C data. Small magnetic moment values even at very fields and near-linear variation of isothermal M below 40 K (Fig. 4) are consistent with such a scenario.

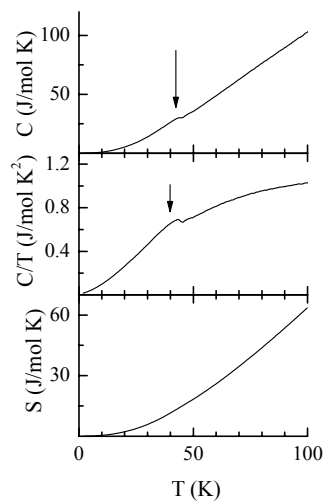

**Fig 3** Heat capacity data for $Ca_3CuRuO_6$, plotted in various ways, along with the total entropy (S). The vertical arrow indicates the ordering temperature ($T_o$).

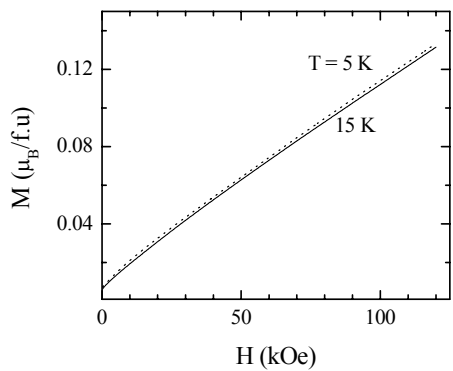

**Fig 4.** Isothermal magnetization behavior for $Ca_3CuRuO_6$.

**CONCLUSION**

We have synthesized stoichiometric $Ca_3CuRuO_6$ and this spin-chain compound apparently undergoes inhomogeneous magnetism below 40 K, presumably due to the formation 'random nano-magnets' arising from crystallographic defects along the chain and geometrical frustration. Considering that the magnetic transition occurs at a temperature far below the magnitude of $\theta_p$, the magnetic frustration resulting from the antiferromagnetic interaction among the magnetic chains arranged in a triangular fashion is clearly operative in this compound.

**REFERENCES**

[1]. K. E. Stitzer et al Curr. Opinion in Solid State Mat. Sci. **5** (2001) 535
[2]. T. N. Nguyen et al J. Solid State Chem. **117** (1995) 300
[3]. Asad Niazi et al Solid State Comm. **120** (2001) 11
[4]. Asad Niazi et al Phys. Rev. Lett. **88** (2002) 107202
[5]. C. A. Moore et al J. Solid State Chem. **153** (2000) 254